\documentclass[%
reprint,
superscriptaddress,
nolongbibliography,
prb,
floatfix,
]{revtex4-2}

\usepackage{graphicx}
\usepackage{dcolumn}
\usepackage{bm}
\usepackage{hyperref}
\hypersetup{
    colorlinks=true,
    linkcolor=blue,
    citecolor=blue,      
    urlcolor=blue,
    }
\usepackage{comment}
\usepackage{soul}
\begin{document}
\title{Spin and current transport in the robust half-metallic magnet $c$-CoFeGe}
\author{Vikrant Chaudhary}
\affiliation{Indian Institute of Technology Roorkee, Department of Chemistry, Roorkee 247667, Uttarakhand, India}
\author{Sapna Singh}
\affiliation{Indian Institute of Technology Roorkee, Department of Chemistry, Roorkee 247667, Uttarakhand, India}
\author{Deepak Gujjar}
\affiliation{Indian Institute of Technology Roorkee, Department of Chemistry, Roorkee 247667, Uttarakhand, India}
\author{Tashi Nautiyal}
\affiliation{Indian Institute of Technology Roorkee, Department of Physics, Roorkee 247667, Uttarakhand, India}
\author{Tulika Maitra}
\affiliation{Indian Institute of Technology Roorkee, Department of Physics, Roorkee 247667, Uttarakhand, India}
\author{Jeroen~van~den~Brink}
\affiliation{Institute for Theoretical Solid State Physics, IFW Dresden, Helmholtzstrasse 20, 01069 Dresden, Germany}
\affiliation{Institute for Theoretical Physics and W\"urzburg-Dresden Cluster of Excellence ct.qmat, Technische Universit\"at Dresden, 01069 Dresden, Germany}
\author{Hem C. Kandpal}\email{Corresponding author: hem.kandpal[at]cy.iitr.ac.in}
\affiliation{Indian Institute of Technology Roorkee, Department of Chemistry, Roorkee 247667, Uttarakhand, India}

\date{\today}
\begin{abstract}
Spintronics is an emerging form of electronics based on the electrons' spin degree of freedom for which  materials with robust half-metallic ferromagnet (HMF) character are very attractive. Here we determine the structural stability, electronic, magnetic, and mechanical properties of the half-Heusler (hH) compound CoFeGe, in particular also in its cubic form. The first-principles calculations suggest that the electronic structure is robust with 100 \% spin polarization at the Fermi level under hydrostatic pressure and uni-axial strain. Both the longitudinal and Hall current polarization are calculated and the longitudinal current polarization ($P_{L}$) is found to be $>99\%$ and extremely robust under uniform pressure and uni-axial strain. The anomalous Hall conductivity (AHC) and Spin Hall conductivity (SHC) of hH cubic CoFeGe (\textit{c}-CoFeGe) are found to be $\sim -100$ S/cm and $\sim 39~\hbar/e$ S/cm, respectively. Moreover, the Curie temperature of the alloy is calculated to be $\sim$524 K with a 3 $\mu_{B}$ magnetic moment. Lastly, the calculated mechanical properties indicate that \textit{c}-CoFeGe is ductile and mechanically stable with a bulk modulus of $\approx$ 154 GPa. Overall, this analysis reveals that cubic CoFeGe is a robust half-metallic ferromagnet and an interesting material for spintronic applications.
\end{abstract}
\maketitle
\section{\label{sec:Intro}Introduction}
The discovery of giant magnetoresistance (GMR) is often regarded as the beginning of spintronics. Spintronic applications focus primarily on information storage, using GMR and tunnelling magnetoresistance (TMR) effects. A typical GMR device is a trilayer structure consisting of ferromagnetic (FM) and nonmagnetic (NM) layers. Up to $20~\%$ GMR was achieved in spin valve systems, devised using NiFeTa and CoFe/Ru/CoFe layered structure and these materials have been used in GMR read head devices \cite{2000Araki}. Materials that can be grown on the commonly used substrates in the layered structure and have robust electronic and magnetic structure, high Curie temperature, and half-metallic behaviour are preferred for such applications. A class of materials showing great promise in this area is the Heusler alloys\cite{2011Graf1HA,2012Casper2HA,2014Wollmann3HA,2017Wollmann4HA,2018Manna5HA}. These alloys are predicted to become half-metal at room temperature. In their electronic structure, one spin channel indicates metallic conduction while the other spin channel is insulating or semiconducting. These half-metal ferromagnets can intrinsically provide single spin channel electrons, with spin polarization reaching unity.

Heusler alloys also gained attention owing to the ability to control their unique electrical and magnetic properties by making changes to their crystalline structures. The Fermi level and band gaps of Heusler alloys can be tuned by substituting other transition elements in parent alloys. Interestingly, it is not difficult to qualitatively predict the behaviour of Heusler alloys from their composition. The magnetic moment of these compounds can be predicted with the help of the Slater-Pauling rule where, the magnetic moment of half and full Heusler (hH and fH) compounds are given as $M=Z-18$ and $M=Z-24$, respectively \cite{2002Galanakis1SP,2006Galanakis2SP,2013Galanakis3SP,2013Galanakis4SP,2013Shaughnessy5SP}. In addition, the HMF behaviour has been studied in detail in NiMnSb \cite{1983Groot1NMS,1988Webster2NMS,2006Galanakis2SP,200Sasloglu54NMS,2001Jenkins5NMS}, FeMnSb \cite{2006Galanakis2SP,2004Mavropoulos1CMS,2013Shaughnessy5SP}, PtMnSb \cite{1983Groot1NMS,1988Webster2NMS}, NiTiSb \cite{2006Galanakis2SP,2013Tabola1CTS,2006Sanyal1NTS}, FeVSb \cite{2017Jianhua1FVS,2006Galanakis2SP} etc. The discovery of NiMnSb with $\approx 99~\%$ spin polarization \cite{2001Jenkins5NMS} is a significant breakthrough establishing that Heusler alloys are potential candidates for spintronic applications. The HMF hH NiMnSb is known for its role as a spin filter for spin-polarized current injection in a semiconductor \cite{2005GalanakisSpinInject}. The HMF Heusler alloys and their significance in spintronic devices have been widely investigated\cite{2021Elphick1Rev,2016Graf2Rev,2016Chris3Rev}. 

Among the known Heusler alloys, the Co-based compounds in particular have received a lot of attention due to their potential in spintronics in the last two decades. A few of the Co-based fH alloys such as Co$_{2}$CrAl \cite{2005Fecher1HCK}, Co$_{2}$MnSi \cite{2006Kandpal2HCK}, and Co$_{2}$FeSi \cite{2006Kandpal2HCK,2005Sabina3HCK,2006Sabina4HCK} are well known and have been thoroughly studied for their electronic and magnetic structures. A device based on Co$_{2}$FeGa$_{0.5}$Ge$_{0.5}$ \cite{2016JungCFGG} was reported to have the largest GMR ($\sim82~\%$) at room temperature. CoMnSb \cite{1988Webster2NMS,2006Galanakis2SP,200Sasloglu54NMS,2004Mavropoulos1CMS}, CoTiSb \cite{2006Galanakis2SP,2013Tabola1CTS,2012Ouardi2CTS}, and CoZrSb \cite{2006Galanakis2SP} are some of the well-explored half Heusler alloys on which many spintronic devices are based. Amongst these, CoMnSb is reported to have $\sim99 \%$ spin polarization and a band gap $\approx$1 eV  in one of the  minority channels. Additionally, a device made of CoFeB was reported to meet the criteria of 1 Gbit magnetoresistive random access memory (MRAM) with a $124~\%$ Tunnel magnetoresistance (TMR) at room temperature \cite{2010Ikeda1CFB}. Much work has been done on CoFeB based devices including some recent works \cite{2021Kunitsyna2CFB,2022Yamamoto3CFB} that have helped in understanding the dynamical spin injection \cite{2022Obinata4CFB}.

In line with the recent studies on Co-based Heusler alloys for spintronics, our focus is the ternary ferromagnetic material CoFeGe with hexagonal Ni$_{2}$In-type crystal structure and a magnetic moment of $2.2$ $\mu_{B}$ per formula unit \cite{1981Szytula1CFG}, where Szytu{\l}a \textit{et al}. observed additional cubic phase signatures along with hexagonal peaks in their X-ray diffraction (XRD) pattern. This presence of cubic phase poses the question on the favourable synthesis conditions and the existence of \textit{c}-CoFeGe. On the other hand, L. Feng \textit{et al.} performed first-principles calculations for many compounds, including CoFeGe in hH structure and investigated the HMF properties \cite{2014Feng2CFG}. However, their work does not address in detail the structural phase transition and spin and current transport in CoFeGe, which is important for spintronics applications. Thus, our interest in Co-based hH alloys as potential spintronic material led us to perform a thorough study of the electronic and anomalous transport and magnetic aspects of CoFeGe. 
\begin{table*}[t]
\caption{\label{tab:table1}
The lattice parameter, magnetic moment, and band gap for CoFeGe in $F\bar{4}3m$ and $P6_{3}/mmc$ space group of the CoFeGe.
}
\begin{ruledtabular}
\begin{tabular}{ccccc}
\textrm{Space Group}&\multicolumn{1}{c}{\textrm{a}}&\multicolumn{1}{c}{\textrm{c/a}}&
\textrm{Magnetic Moment}&\textrm{Minority band gap }\\
\textrm{ }&
\multicolumn{1}{c}{\textrm{(\AA)}}&
\multicolumn{1}{c}{}&
\textrm{($\mu_B$/f.u.)}&\textrm{($E_g^{\downarrow}$, eV)}\\
\colrule
\rule{0pt}{3ex}
$F\overline{4}3m$ & 5.4967 & 1 & 3.0000 &0.4399\\
$P6_3/mmc$ & 4.0821 & 1.2352 & 2.8862 &0.00\\
\end{tabular}
\end{ruledtabular}
\end{table*}

In this work, utilizing the \textit{ab-initio} approach, we systematically investigated the structural, electronic, and magnetic properties of CoFeGe. These properties were also studied under external pressure and uni-axial strain. The uni-axial strain breaks the cubic symmetry and CoFeGe adapts the tetragonal crystal structure. The static, dynamic, and mechanical stability of CoFeGe was investigated and the thermodynamic stability was confirmed from data available at AFLOW \cite{AFLOW} and OQMD \cite{1OQMD,2OQMD}.  The spin polarization from the density of states ($P_{DOS}$), longitudinal current polarization ($P_{L}$), Hall current polarization ($P_{Hall}$), anomalous Hall conductivity (AHC), and spin Hall conductivity (SHC) have also been evaluated. Further, the mechanical properties have also been reported from first principles calculations.

\section{\label{sec:Comp}COMPUTATIONAL DETAILS}
We have made use of the first-principles density functional theory (DFT) under the projector augmented wave (PAW) formalism based Vienna \textit{ab-initio} Simulation Package (VASP) \cite{1VASP,2VASP,3VASP,4VASP,5VASP} within the generalized gradient approximation (GGA) developed by J. P. Perdew, K. Burke, and M. Ernzerhof (PBE) \cite{GGA}. The crystal structure was optimized on a $41\times 41\times 41$ k-mesh using 550 eV  cutoff energy and with a charge density convergence criteria of $1.0\times 10^{-7}$. The energy convergence criteria in the self-consistent field (SCF) cycles was set to $1.0 \times 10^{-8}$ eV. The phonon band structures were obtained with the help of Phonopy package \cite{phonopy} interfacing it with VASP. We also used VASPKIT \cite{VASPKIT} for various pre-processing and post-processing tasks for VASP calculations.

The Wannier90 package \cite{Wann90} was used with VASP to generate maximally localized Wannier functions (MLWFs). These MLWFs were generated on a $9\times9\times9$ DFT k-mesh and were converged up to an order of $10^{-11}$ and $10^{-10}$ during disentanglement and spread calculations, respectively. We obtained a total spread of 37.46 \AA$^{2}$. These well-converged and localized MLWFs were further used to calculate AHC and SHC on a $200\times200\times200$ Berry mesh with an adaptive refinement of $7\times7\times7$ whenever Berry curvature exceeded 100 \AA$^{2}$. 

The electrical transport properties at different pressure and strain conditions were calculated using Boltzmann theory and relaxation time approximation as implemented in the Boltztrap code  \cite{BoltzTraP2}. The electrical conductivity has been calculated with respect to relaxation time ($\tau$) which was further used in the evaluation of the $P_{L}$.

\section{\label{sec:Res}Results and Discussion}
\subsection{\label{subsec:Struct}Structural Analysis}
\begin{figure}[t]
\includegraphics[scale=0.8]{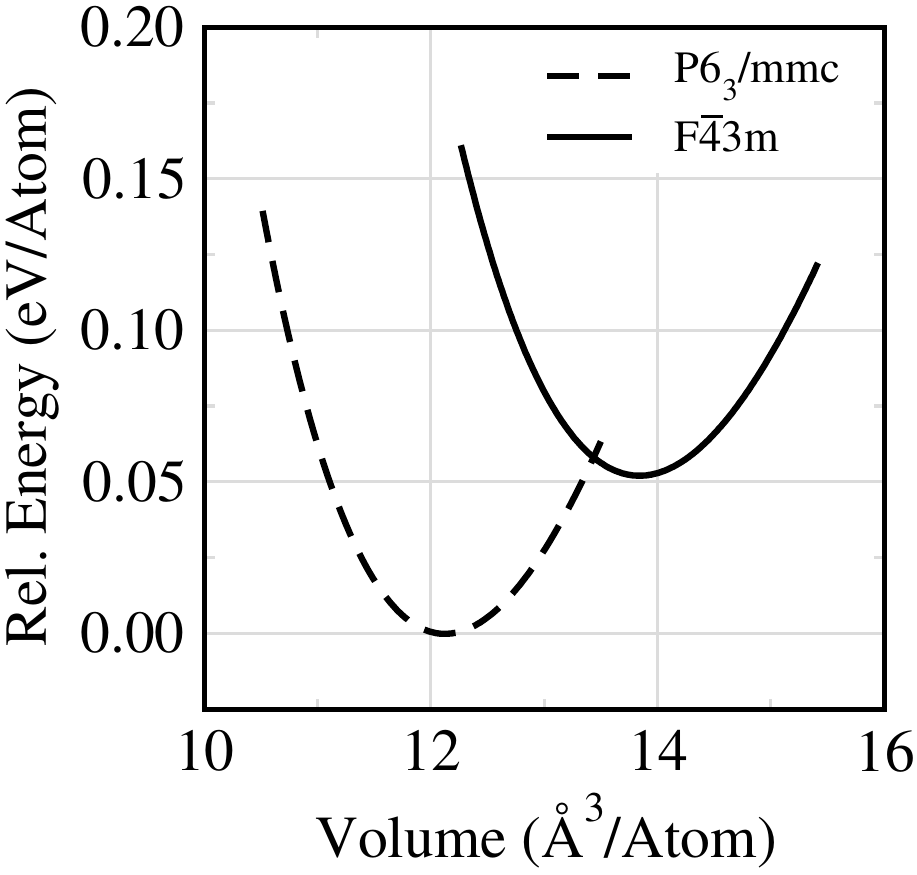}
\caption{\label{fig:eos}The calculated energy per atom on a relative scale as a function of volume per atom for cubic and hexagonal CoFeGe.}
\end{figure}

As discussed in section \ref{sec:Intro}, the ternary inter-metallic CoFeGe exists in the hexagonal phase, space group $P6_{3}/mmc$ (No. 194), with a signature of minor cubic phase in the XRD pattern. To understand the possibility of the synthesis of cubic phase, space group $F\bar{4}3m$ (No. 216), we investigated the static, dynamic, and thermodynamic stability of CoFeGe as discussed ahead.

The static stability requires structure optimization on a sufficiently dense k-mesh, thus, we have used a fine k-mesh as discussed in section \ref{sec:Comp}. The energy vs volume optimization was done by calculating energies in the volume range from $-10~\%$ to $10~\%$ in a step of $2~\%$ change. In order to obtain the optimized lattice parameters, energy vs volume data was fitted with the Birch-Murnaghan (BM) equation of state \cite{1BM,2BM,3BM} as shown in Fig. \ref{fig:eos}. All the non-magnetic (NM), ferromagnetic (FM), and anti-ferromagnetic (AFM) phases of both the structures were optimized and it was found that the FM phase is favourable over the other two for both the structures. Thus, only the FM phase of both the structures are shown in Fig. \ref{fig:eos}. Table \ref{tab:table1} lists the lattice parameters, magnetic moment, and band gap of both the structures. The ground state energy difference (per atom) between hexagonal and \textit{c}-CoFeGe is 50 meV, which indicates that the hexagonal phase is energetically favourable at ambient conditions. However, this small energy difference leads to chances of achieving cubic phase \textit{e. g.} by inducing internal pressure in the system by methods such as substituting larger size atoms or synthesis on a substrate etc. These methods can lead to an increase in  the volume of the system, making the cubic phase energetically favourable over the hexagonal phase (Fig.\ref{fig:eos}). 
\begin{figure}[ht]
    \centering
\begin{tabular}[t]{lc}
    \includegraphics[scale=0.7]{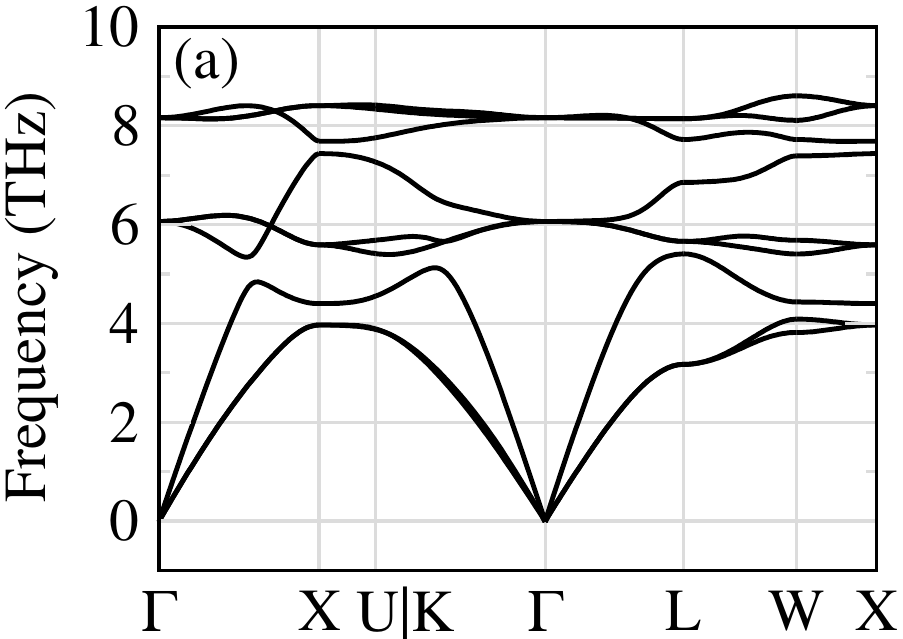}\\
    \includegraphics[scale=0.73]{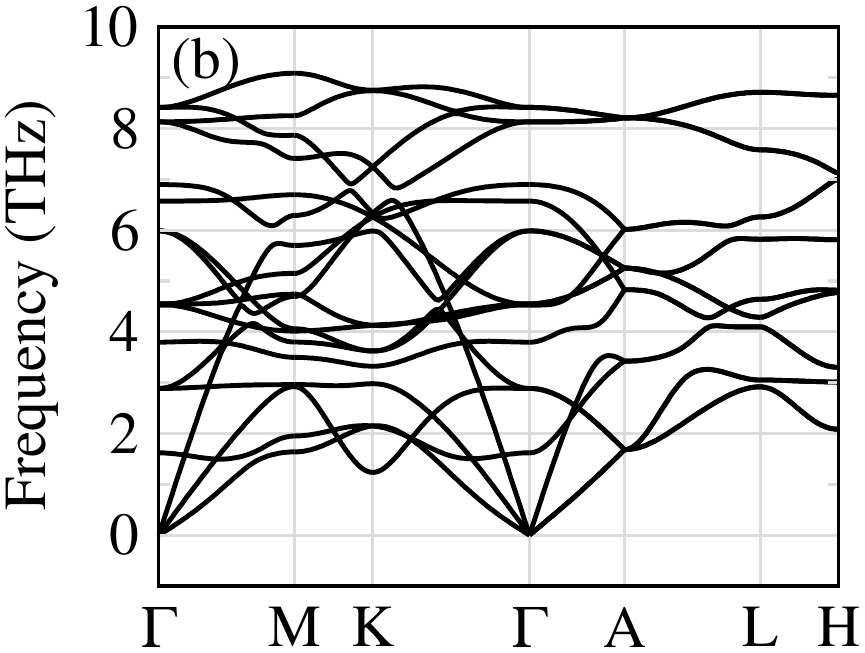}
\end{tabular}
\caption{\label{fig:dis}The phonon band structure of (a) the $F\bar{4}3m$ and (b) $P6_{3}/mmc$ phases of CoFeGe.}
\end{figure}

Next, we checked for the dynamical stability of  the cubic and the hexagonal analogues of CoFeGe with the help of phonon dispersion curves. The phonon calculations are performed in accordance with the density functional perturbation theory (DFPT) implemented in VASP and Phonopy. For the cubic structure, a $2\times 2\times 2$ supercell was generated and the calculations were performed on a $5\times5\times5$ k-mesh in the phonon Brillouin zone. Similarly, for the \textit{h}-CoFeGe, we used a k-mesh $4\times4\times5$ for a $3\times 3\times 2$ supercell. The force constants were extracted from these calculations using Phonopy and were used for obtaining the phonon dispersion curves. Phonons are basically the normal modes or quanta of vibrations in a crystal which help in governing the stability of a system. For a system to be dynamically stable, it should have only real phonon frequencies and not imaginary ones. Fig. \ref{fig:dis} (a) and (b) show 9 (3 acoustical and 6 optical) and 18 (3 acoustical and 15 optical) phonon branches of the cubic and the hexagonal structures, respectively, of CoFeGe. The absence of the imaginary modes asserts the dynamic stability of both the structures.

In addition to static and dynamic stabilities, thermodynamic stability is also important for predicting new materials. Interestingly the $P6_{3}/mmc$ structure of CoFeGe is thermodynamically unstable according to the data available at OQMD \cite{1OQMD,2OQMD} and AFLOW \cite{AFLOW}, whereas it has been realized experimentally \cite{1981Szytula1CFG}. Similarly, there are many thermodynamically unstable materials which have been successfully synthesized by experimental groups. For example, the fH Ni$_{2}$CuSn lies $\approx$ 32 meV/Atom above the convex hull (available at OQMD) but it has been experimentally realized and studied extensively \cite{1N2CS, 2N2CS,3N2CS}. Another example is the W doped Fe$_{2}$VAl, which exhibited the best thermoelectric performance \cite{F2VA}. The Fe$_{2}$V$_{0.8}$W$_{0.2}$Al was found metastable through the DFT calculations but was successfully synthesized in the form of thin film Heusler alloy. Cu$_{3}$N and the bulk ${\rm{NdNi}}{{\rm{O}}_2}$ are also the examples of metastable and unstable materials, of which Cu$_{3}$N \cite{TF1} was synthesized and a significant reduction was achieved in the instability of  ${\rm{NdNi}}{{\rm{O}}_2}$ \cite{NNO} was achieved.  ${\rm{NdNi}}{{\rm{O}}_2}$ has an energy +176 meV/atom \cite{NNO}, while \textit{c}-CoFeGe is +120 meV/atom above the hull as per the data available on OQMD. 

\begin{figure}[t]
\includegraphics[scale=0.9]{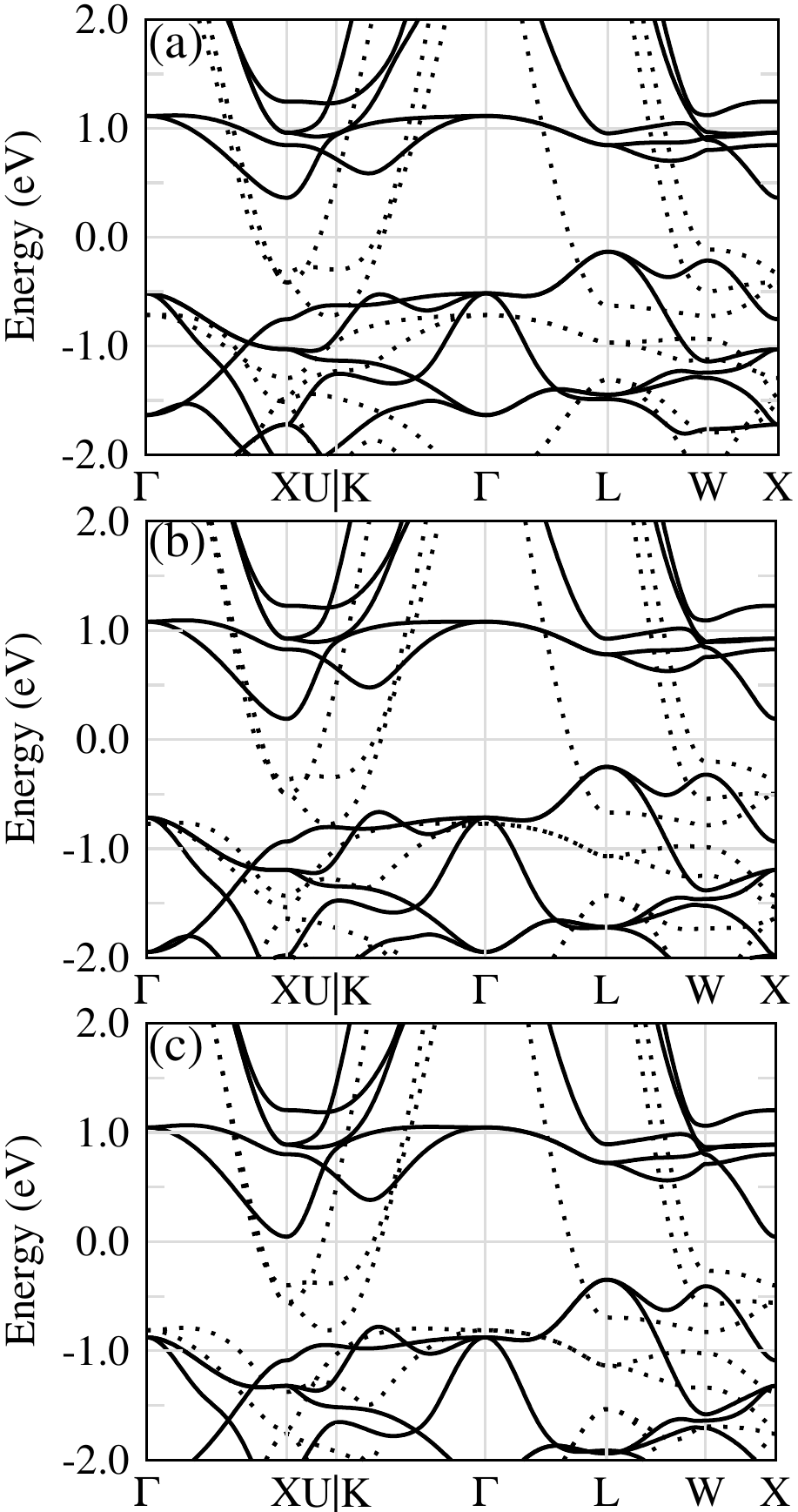}
\caption{\label{fig:cbands}The bandstructure of \textit{c}-CoFeGe at (a) -10 GPa, (b) ambient, and (c) 10 GPa pressure. The solid and dotted curves are spin-down and spin-up bands, respectively.}
\end{figure}
\begin{figure}[t]
\includegraphics[scale=0.85]{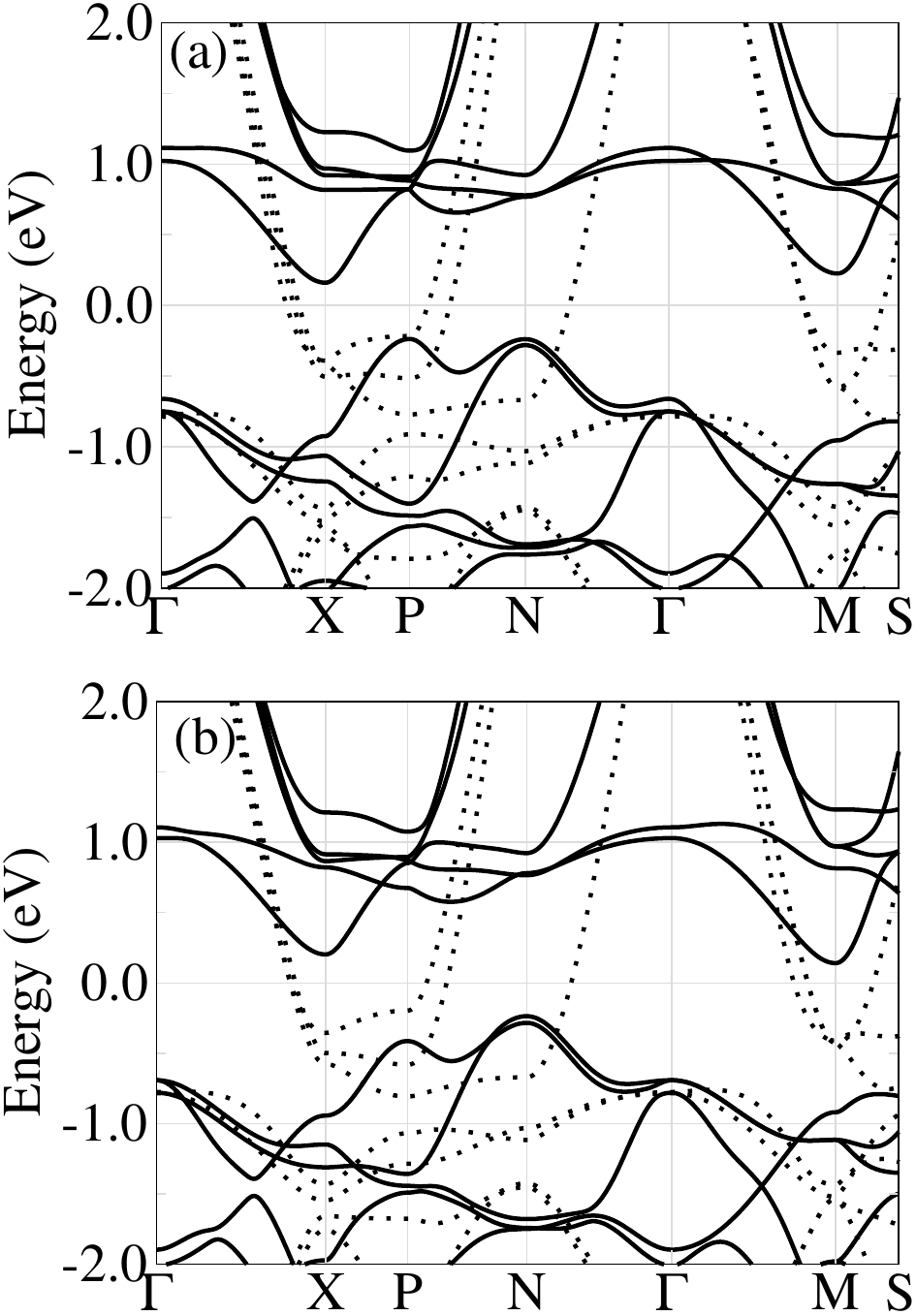}
\caption{\label{fig:tbands}The bandstructure of the tetragonal CoFeGe, when (a) $c = 0.98a_{0}$ and (b) $c = 1.02a_{0}$, where $a_{0}$ is the lattice constant of the relaxed \textit{c}-CoFeGe.}
\end{figure}
These findings suggest that though the thermodynamical stability gives an idea about the structural stability of the system, it does not conclusively predict the possibility of synthesis of a material. The formation energy of hH CoFeGe is negative (data available on OQMD) with respect to the constituting elements. On considering other binary and ternary decompositions, the formation energy is -67 meV/Atom and the energy above the Hull is 120 meV/Atom. The hexagonal phase has -123 meV/Atom formation energy and 64 meV/Atom energy above the convex Hull, as per the data available on OQMD. Many experimentally known thermodynamically metastable compounds have less than 100 meV/atom energy above the convex hull. There are also some experimentally known compounds that have more than 100 meV/atom energy above the hull \cite{metastable}. Since both \textit{h}-CoFeGe and \textit{c}-CoFeGe are unstable as per first-principles data on OQMD, but the \textit{h}-CoFeGe does exist experimentally, hence even the cubic structure may be realized experimentally.

\subsection{\label{subsec:elec}Electronic and Magnetic Properties}
\begin{figure*}[ht]
\includegraphics[scale=0.9]{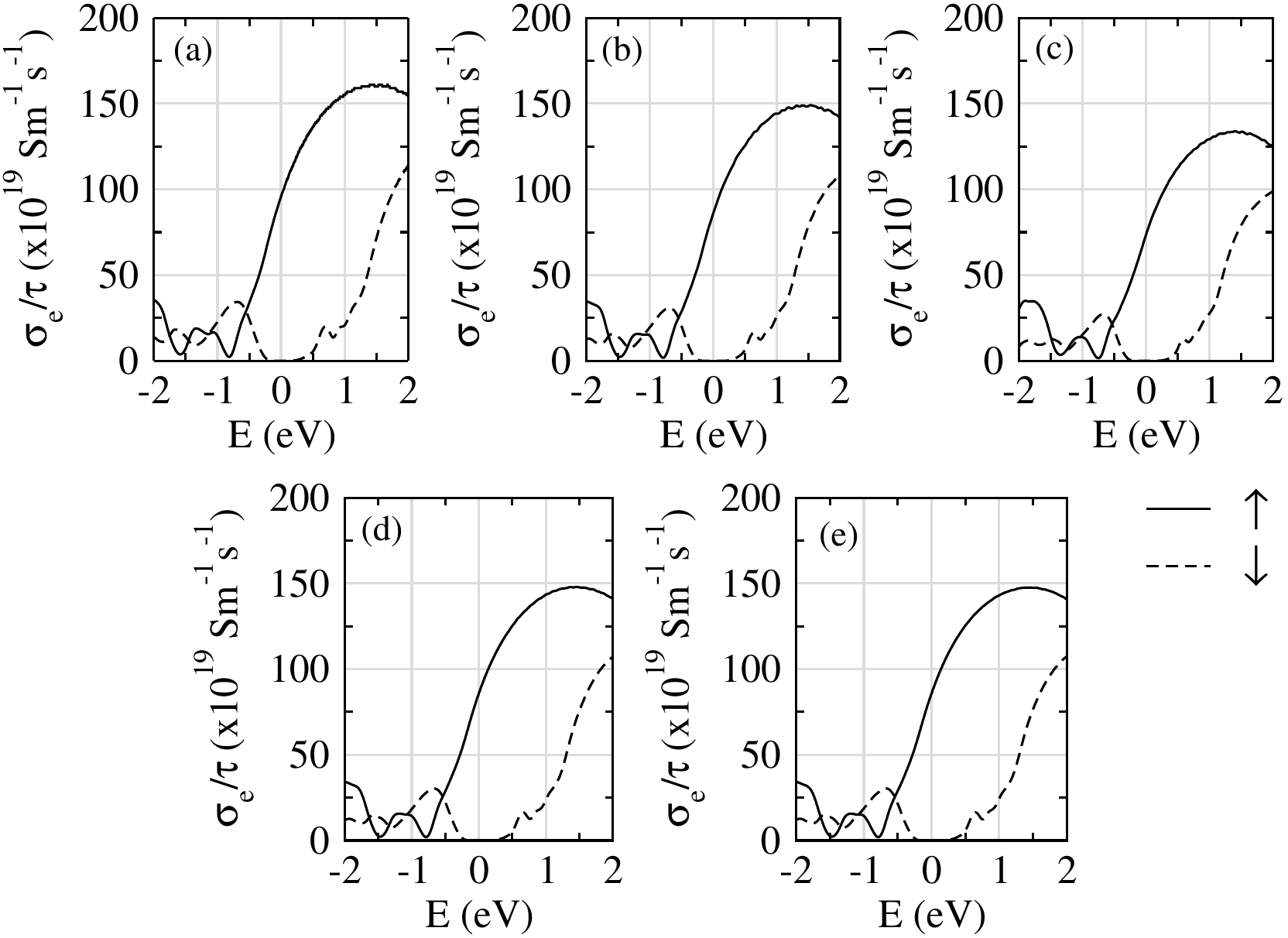}
\caption{\label{fig:P_Long} The spin decomposed electrical conductivity of \textit{c}-CoFeGe at (a) 10 GPa, (b) ambient, and (c) -10 GPa pressure. Fig. (d) and (e) show the electrical conductivity under $2~\%$ compressive and equally elongative strain, respectively, along the \textit{c}-axis. The solid (dotted) curve represents the spin-up (spin-down) component of the electrical conductivity and the Fermi level has been shifted to 0 eV.}
\end{figure*}

The electronic band structure of HMF \textit{c}-CoFeGe was calculated using GGA-PBE approximation as implemented in the VASP code. As discussed in section \ref{subsec:Struct}, the ground state is FM for both ($F\bar{4}3m$ and $P6_{3}/mmc$) structures. The cubic structure is half metallic, whereas the hexagonal is metallic as both spin (up and down) bands are present at the Fermi level. This confirms the HMF nature of CoFeGe important for applications in spintronics. Next, we also checked for the robustness of this HMF character. Also, an impact of negative pressure was observed in the alloy for a better understanding of the durability under extreme conditions. A uniform pressure was applied to the \textit{c}-CoFeGe and the half-metallic behaviour was found to remains intact within a pressure range of -10 GPa to 10 GPa as shown in Fig. \ref{fig:cbands}. In order to further support the robustness of the HMF behaviour, the electronic structure was studied under a compressive as well as elongative strain of $2~\%$ along the c-direction keeping the volume fixed. As shown in Fig. \ref{fig:tbands}, the CoFeGe remains an HMF with a $\approx$ 0.43 eV band gap in the minority channel. To conclude, our electronic structure investigations suggest that the HMF behaviour of \textit{c}-CoFeGe is extremely robust, and the spin polarization remains $100~\%$ at the Fermi level, which was confirmed with the help of the equation,
\begin{equation}
 P_{DOS}=\left(\frac{N_{\uparrow}-N_{\downarrow}}{N_{\uparrow}+N_{\downarrow}}\right)_{E_{F}},
\end{equation}
where $N_{\uparrow}$ and $N_{\downarrow}$ are spin-up and spin-down electronic states present at the Fermi level.

In most of the experiments, currents are studied and the $P_{DOS}$ may not give us the important information required for the spintronic applications. Therefore, we also calculated the longitudinal current polarization ($P_L$) and the Hall current polarization ($P_{Hall}$). For $P_{L}$ calculation, we need electrical conductivity of spin up ($\sigma_{\uparrow}$) and spin down ($\sigma_{\downarrow}$) states and corresponding relaxation times ($\tau_{\uparrow}$ and $\tau_{\downarrow}$) as given by
\begin{equation} P_{L}=\frac{\sigma_{\uparrow}/\tau_{\uparrow}-\sigma_{\downarrow}/\tau_{\downarrow}}{\sigma_{\uparrow}/\tau_{\uparrow}+\sigma_{\downarrow}/\tau_{\downarrow}}.
\end{equation}

Since it is challenging to calculate $\tau_{\uparrow}$ and $\tau_{\downarrow}$, we used an approximation that the relaxation time does not depend on the k-points, energy, and the direction of spin. This assumption allows us to use same relaxation time ($\tau_{\uparrow}$ = $\tau_{\downarrow}$) for both spins. We used BoltzTrap2 \cite{BoltzTraP2} code based on semi-classical Boltzmann transport theory for the calculation of longitudinal electrical conductivity ($\sigma_{\uparrow}$ and $\sigma_{\downarrow}$) divided by the corresponding relaxation times. We employed a $41\times41\times41$ k-mesh for obtaining ground state charge density and energies and used those as input in BoltzTrap2 code for further calculations.  

Fig. \ref{fig:P_Long} shows that the trends in spin decomposed electrical conductivity at -10 GPa, 10 GPa, and after uni-axial strain match well with the electrical conductivity at the ambient pressure. The electrical conductivity below the valance band maximum (VBM) is low in both spin channels; as soon as VBM is crossed, $\sigma_{\uparrow}$ increases rapidly while $\sigma_{\downarrow}$ picks up slowly and approaches a negligible value with respect to the $\sigma_{\uparrow}$. Our results show that the $P_{L}$ is $>99~\%$ at the Fermi level in all the cases. Above the conduction band minimum, the $\sigma_{\downarrow}$ also increases rapidly. 

Another crucial property from the spintronic viewpoint is the Hall current (spin) polarization given as \cite{1PH,2PH}
\begin{equation}
 P_{Hall}=\frac{\sigma_{xy}^{HC\uparrow}-\sigma_{xy}^{HC\downarrow}}{\sigma_{xy}^{HC\uparrow}+\sigma_{xy}^{HC\downarrow}}.
\end{equation}
The $\sigma_{xy}^{HC\uparrow}$ ($\sigma_{xy}^{HC\downarrow}$) is the spin up (down) Hall conductivity and can be obtained with the help of AHC and SHC. The simplified expression for $P_{Hall}$ is
\begin{equation}
 P_{Hall}=\frac{2e}{\hbar}\frac{\sigma_{xy}^{SHC}}{\sigma_{xy}^{AHC}},
\end{equation}
where $\sigma_{xy}^{SHC}$ and $\sigma_{xy}^{AHC}$
are spin Hall conductivity (SHC) and anomalous Hall conductivity (AHC), respectively \cite{1CSP,2CSP}. 

The AHC has been studied extensively for the Heusler class and the fH compounds show large AHC (e. g. Co$_2$MnAl $\sim1420$ S/cm) \cite{Yimin_2022}, arising due to the presence of Weyl points, nodal lines, and band crossings near the Fermi level \cite{1CFAHC,1FHAHC}. On the other hand, the hH alloys show a relatively lower AHC value. The well-known hH alloys, GdPtBi \cite{RPtBi}, TbPtBi \cite{TbPtBi} and HoPtBi \cite{HoPtBi} are reported to have an AHC close to 60 S/cm, 100 S/cm and -75 S/cm, respectively. The AHC in topological semimetal TbPtBi was tuned up to $\approx$ 125 S/cm with the help of magnetic field and temperature. 
\begin{figure}[t]
\includegraphics[scale=0.7]{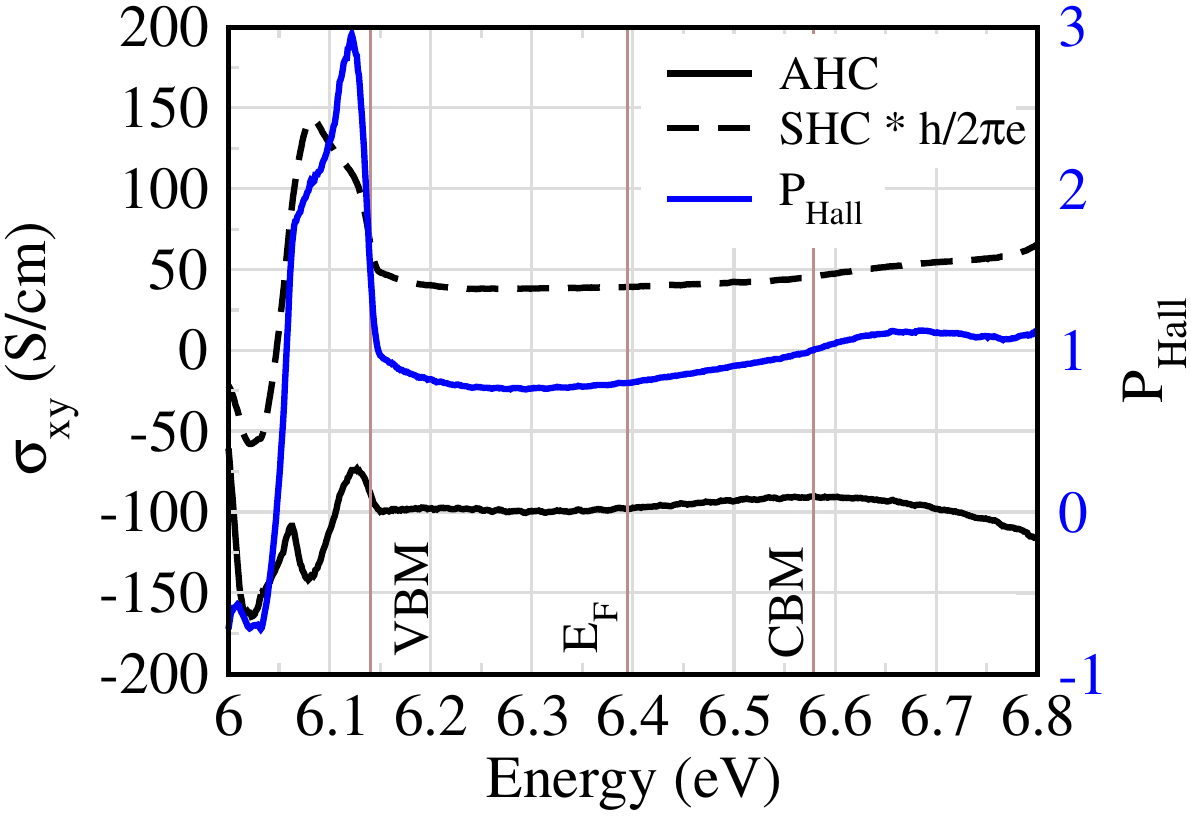}
\caption{\label{fig:ahc}The AHC, SHC, and $P_{Hall}$ for \textit{c}-CoFeGe. The VBM, CBM, and Fermi level ($E_F$) are indicated using vertical lines.}
\end{figure}

Our calculations show that in comparison to many popular hH alloys discussed above, the hH CoFeGe has a large AHC ($\approx$ 100 S/cm) in the minority channel band gap, as shown in Fig.\ref{fig:ahc}. The largest value of AHC and SHC is found to be -164.67 S/cm at 0.37 eV and 143.23 $e/\hbar$ S/cm at 0.31 eV below the Fermi level, respectively. The AHC and SHC change rapidly below the VBM and become almost constant within the energy gap, as shown in Fig. \ref{fig:ahc}, resulting in an absolute $P_{Hall}$ value between 0.79 and 1.00. The maximum of AHC and SHC occurs a little below the VBM possibly resulting from the many band crossings seen at/around -0.37 eV and -0.31 eV as shown in Fig. \ref{fig:cbands}(b). These crossings split up after switching on the spin-orbit coupling, leading to large Berry curvature and a large AHC and SHC. The current (spin), static DOS polarization and AHC calculations suggest that the \textit{c}-CoFeGe may be a promising candidate for spintronic devices. The magnetic structure is also of importance in the study of spin transport and is discussed in the following section.

The Curie temperature (T$_{C}$) of \textit{c}-CoFeGe was calculated within the mean field approximation (MFA) \cite{Curie} using the spin-polarized relativistic Korringa-Kohn-Rostoker (SPR-KKR) code \cite{1SPRKKR,2SPRKKR}. The predicted value of T$_{C}$ as 528 K for \textit{c}-CoFeGe is quite large when compared with the measured value of 370 K \cite{1981Szytula1CFG} for \textit{h}-CoFeGe. We found the Curie temperature of \textit{c}-CoFeGe as $\approx$ 528 K with a magnetic moment of 3 $ \mu_{B}$ per formula unit. The Fe and Co atoms contribute approximately 2.5 $\mu_{B}$ and 0.5 $\mu_{B}$ to the total magnetic moment, while the contribution of Ge is close to zero. The value of $T_{C}$ as 528 K for \textit{c}-CoFeGe is comparable with that for other well-known hH alloys, i.e. PtMnSb (582K)\cite{1988Webster2NMS}, CoMnSb (490K)\cite{1988Webster2NMS,200Sasloglu54NMS}, NiTiSb (330 K)\cite{2006Sanyal1NTS} etc. The fH alloys show relatively higher $T_{C}$, for example, the $T_{C}$ and magnetic moment of Co$_{2}$FeSi have been experimentally reported to be 1100 K and 6 $\mu_{B}$ \cite{4HCK}, supposed to be the highest known values for a Heusler HMF alloy. In general, the Curie temperature of the Heusler alloys mostly falls between 200 K and 1200 K. Hence \textit{c}-CoFeGe is predicted to have a high $T_{C}$, making it a promising candidate in spin transport applications. 

\subsection{Mechanical Properties}
Having investigated \textit{c}-CoFeGe for applications in spin transport, it is worthwhile to explore its mechanical properties. Heusler alloys are generally ductile and their mechanical properties are compiled in a review article \cite{MchanicalH}.
\begin{table}[t]
\caption{\label{tab:table3}%
Mechanical properties of hH CoTiSb and hH CoFeGe. The modulus, hardness, and Cauchy pressure are in GPa, whereas the Poisson's ratio and Pugh's ration are dimensionless. The experimentally measured quantities are in paranthesis.}
\begin{ruledtabular}
\begin{tabular}{lll}
Mechanical Properties & hH CoTiSb \cite{TCS,2012Ouardi2CTS} & hH CoFeGe\\
\colrule
Bulk Modulus (B)& 142 (166)  & 154 \\
Young's Modulus (Y)& 224  & 148 \\
Shear Modulus (G)& 91  & 55 \\
Poisson's Ration (v)& 0.24 & 0.34\\
Vickers Hardness (V)& & 4.89 \\
Pugh's Ratio (B/G)& 1.56 & 2.80\\
Cauchy Pressure (P$_C$)& & 62.7 \\
\end{tabular}
\end{ruledtabular}
\end{table}

For a better insight into this study, we used hH alloy TiCoSb as a reference system in order to adjudge our results as shown in Table \ref{tab:table3}. The bulk modulus of TiCoSb has been experimentally measured to be $\approx$ 166 GPa with no structural phase transition up to 115 GPa external pressure \cite{TCS}.  The calculated bulk modulus (154 GPa) of \textit{c}-CoFeGe is close to the bulk modulus of hH TiCoSb, suggesting that a large external pressure would be needed for the structural phase transition. The calculated Young's modulus ($\approx$ 148 GPa) of \textit{c}-CoFeGe is also fairly high (from Voigt \cite{Vogit} methods). 

The calculated value of Pugh's ratio, one of the important properties to understand the mechanical nature of the materials, is $2.80$ for \textit{c}-CoFeGe. This value is $>1.75$, indicating that the compound is ductile. The stiffness tensor of any cubic structure has primarily three mechanical constants; C11, C12, and C44. These constants are used to calculate the mechanical properties and understand the mechanical stability \cite{MechStability}. The \textit{c}-CoFeGe meets the elastic stability criteria ($C11 - C12 > 0$, $C11 + 2C12 > 0$, and $C44 > 0$), making it mechanically stable.
\section{Conclusions}
We have thoroughly studied CoFeGe from spintronic viewpoint. The hexagonal phase of CoFeGe is already known but the signature of a cubic phase in the XRD pattern of the compound prompted us to check the possibility of structural phase transition from hexagonal ($P6_{3/mmc}$) to cubic ($F\bar{4}3m$) analogue. The small energy difference of the order of  $\approx$ 51 eV/Atom between the hexagonal and cubic phases can be overcome to realize the cubic phase. In its ground state, the \textit{c}-CoFeGe is ferromagnetic with half-metallic behaviour. The HMF character is preserved with 100 \% spin polarization within extreme condition of the pressure range of -10 GPa to 10 GPa, and under $2~\%$  compressive and elongative strain. Further, the longitudinal and Hall current spin polarization show promising values with the $P_{L}$ staying $>99~\%$ spin-polarized and $P_{Hall}$ value changes from $80~\%$ to $100~\%$ between the VBM and CBM of minority spin channel. The calculated $T_{C}$ and magnetic moment of \textit{c}-CoFeGe are $\approx$ 524 K and 3 $\mu_{B}$, respectively. Lastly, we investigated the mechanical properties of the \textit{c}-CoFeGe and observed that the structure is ductile and mechanically stable. Thus \textit{c}-CoFeGe is predicted to be a robust HMF with large $P_{L}$ and moderately high T$_{C}$ values. In nutshell, \textit{c}-CoFeGe, when realized in hH structure, will be an interesting candidate for spin transport applications.

\begin{acknowledgments}
This work used the Supercomputing facility of IIT Roorkee established under National Supercomputing Mission (NSM), Government of India and supported by Centre for Development of Advanced Computing (CDAC), Pune. We have also used other computational facilities provided by Institute Computer Center (ICC), IIT Roorkee. VC wish to acknowledge the financial support received from Ministry of Education, Government of India.
\end{acknowledgments}

\end{document}